\documentclass[12pt]{article}
\usepackage{graphics, color}
\usepackage{graphicx}
\usepackage{amssymb}

\def\gsim{\, \rlap{$>$}{\lower 1.1ex\hbox{$\sim$}}\,}
\def\lsim{\, \rlap{$<$}{\lower 1.1ex\hbox{$\sim$}}\,}
\newcommand{\pho}{F}

\begin{document}


\begin{titlepage}
\bigskip
\bigskip\bigskip\bigskip
\centerline{\Large A Matrix Model for Black Hole Thermalization}
\bigskip\bigskip\bigskip

\centerline{{\bf Norihiro Iizuka}\footnote{\tt iizuka@kitp.ucsb.edu}}
\medskip
 \centerline{{\bf Joseph Polchinski}\footnote{\tt joep@kitp.ucsb.edu}}
\medskip
\centerline{\em Kavli Institute for Theoretical Physics}
\centerline{\em University of California}
\centerline{\em Santa Barbara, CA 93106-4030}\bigskip
\bigskip\bigskip
\bigskip\bigskip


\begin{abstract}
We present a matrix model which is intended as a toy model of the gauge dual of an AdS black hole.  In particular, it captures the key property that at infinite $N$ correlators decay to zero on long time scales, while at finite $N$ this cannot happen.  The model consists of a harmonic oscillator in the adjoint 
which acts as a heat bath for a particle in the fundamental representation.  The Schwinger-Dyson equation reduces to a closed recursion relation, which we study by various analytical and numerical methods.  We discuss some implications for the information problem.
\end{abstract}
\end{titlepage}
\baselineskip = 16pt

\setcounter{footnote}{0}


\section{Introduction}

The black hole information paradox~\cite{{Hawking:1976ra}} is one of the great thought experiments in physics.  Three decades of effort have made it clear that it has no trivial resolution, but indeed requires a modification of some central principle of physics~\cite{Hawking:1976ra}, although the focus has largely shifted from a breakdown of quantum purity to a holographic nonlocality of quantum gravity~\cite{'t Hooft:1993gx,Susskind:1994vu}.

The information paradox was largely responsible for the intense scrutiny given to the dynamical properties of black branes, which led to the discovery of gauge/gravity duality~\cite{Maldacena:1997re}.  This duality in turn implies that information is not lost, because one can describe the formation and decay of a black hole within systems that have a well-defined dual description in an ordinary quantum framework.\footnote{In stating that the dual theory is well defined, we have in mind the superrenormalizable duals introduced in Ref.~\cite{{Itzhaki:1998dd}}, as well as the BFSS Matrix Theory~\cite{Banks:1996vh}, which is a quantum mechanical example of gauge/gravity duality.}

This argument for information preservation is rather indirect.  In order to calculate the black hole S-matrix, one must translate the initial infalling state into the dual field theory, evolve forward in the field theory variables, and translate back into the outgoing state of the Hawking radiation.  It would be desirable to have a prescription entirely in terms of the bulk gravitational variables, since this is how the Hawking radiation, and the apparent information loss, is found.  Indeed, the proposed answers to the question ``Where does the argument for information loss break down?'' are now as diverse as the answers to the original question ``What happens to information thrown into a black hole?''  We may hope that the attempt to answer this new question will be as fruitful as it was for the earlier one.

It is notable that some features of the black hole persist even in the weakly coupled gauge theory, where the spacetime interpretation of the bulk breaks down.  The continuation of the Hawking-Page transition~\cite{Hawking:1982dh, Witten:1998zw} to weak coupling has been studied extensively~\cite{Skagerstam:1983gv,Sundborg:1999ue,Polyakov:2001af,Aharony:2003sx}.  It has also been argued that vestiges of the black hole singularity~\cite{Festuccia:2005pi} and of the information problem~\cite{Festuccia:2006sa} survive at weak coupling.\footnote{Ref.~\cite{Festuccia:2005pi}, and our work, model the weakly coupled duals to the $AdS_5 \times S^5$ or Matrix Theory black holes.  The weakly coupled limit of the BTZ black hole is also interesting, and the corresponding questions have been explored in Refs.~\cite{Maldacena:2001kr,Birmingham:2001pj,Birmingham:2002ph,Balasubramanian:2005qu}.}

Ref.~\cite{Festuccia:2006sa} considers the information problem in the form presented in Ref.~\cite{Maldacena:2001kr}.  The low energy gravitational field theory description of AdS black holes shows quasinormal behavior, the exponential decay of correlations in time~\cite{Horowitz:1999jd,Danielsson:1999fa}.  This description should be valid at large $N$ and large 't Hooft parameter, where the curvature is small. In the dual field theory, exponential decay is possible at infinite $N$, where the thermal field theory has an infinite number of states and can absorb an arbitrary amount of information.  However, at large finite $N$ exponential decay can only persist until the correlations are of order $e^{-O(N^2)}$, because the black hole has only a finite number of states and so the correlator is a sum of a finite number of exponentials. Reconciling this with the prediction of the low energy bulk field theory is a manifestation of the information paradox.\footnote{The relation between the information paradox and the discreteness of the spectrum is discussed further in Ref.~\cite{Balasubramanian:2006iw}. }

Ref.~\cite{Festuccia:2006sa} argues that signs of the quasinormal behavior can be seen at arbitrarily weak coupling, in that perturbation theory breaks down at long times no matter how small the coupling is.  We would like to extend this, resumming the perturbation theory to see the explicit form of the late-time behavior.  The graphs considered in Ref.~\cite{Festuccia:2006sa} have a simple iterative structure that suggests such a resummation.  However, we would like to find a situation in which this resummation is systematic, in that it represents the full planar amplitude.  This would provide a setting for discussing the breakdown of the large-$N$ approximation, which is dual to the loop expansion of the bulk theory.

We identify a simple system with the desired property: a harmonic oscillator in the $U(N)$ adjoint representation plus a harmonic oscillator in the $U(N)$ fundamental, coupled through a trilinear interaction.  In particular, it is sufficient to consider the limit of a single fundamental excitation in interaction with the adjoint oscillator.  One can think of the adjoint as a heat bath coupled to the fundamental, and we study the decay of correlations of the fundamental field.  

This model may have a number of realizations; one can think of it as a reduced version of a D0-brane black hole with a D$0$ probe, where the matrix variables are the 0-0 fields and the fundamentals are the 0-$0_{\rm probe}$ fields~\cite{Iizuka:2001cw,Iizuka:2002wa}.  It might actually arise in some decoupling limit, although of course it would be a limit of large spacetime curvature.  This simple model may also have applications outside of black hole physics, along the lines of the Caldeira-Leggett model~\cite{{Caldeira:1982iu}}.

In Sec.~2 we introduce the model.  It has the same graphical structure as the 't Hooft model of two-dimensional QCD~\cite{'t Hooft:1974hx}.  Unlike the 't Hooft model there are dynamical adjoints, but we achieve the simplified graphical structure of the 't Hooft model by stipulating that the adjoints have no self-interaction.  We first consider zero temperature; we derive the Schwinger-Dyson equation, which reduces to a two-term recursion equation with respect to frequency.  The singularities consist of poles on the real axis.  We also analyze the quantum mechanics canonically, leading to a closed-form solution a zero temperature.

In Sec.~3 we consider nonzero temperature.  Again, the Schwinger-Dyson equation reduces to an iterative equation, but with three terms rather than two.  In this case we can argue that the correlator for the fundamental field must approach zero at long times (in the planar limit).  Numerically, we show that the decay is power law for small couplings and exponential for couplings that are sufficiently large; the power law regime is likely an artifact of our model that has no analog for the black hole.  We develop briefly the approximate solutions for small coupling and small mass.  We also extend the model, by the introduction of decoupled sectors and the singlet constraint, so as to obtain a Hagedorn transition.

In Sec.~4 we discuss some implications for the information problem.

\section{Zero temperature}

\subsection{The model}

The fields are a Hermitian matrix  $X_{ij}(t)$ and a complex vector $\phi_i(t)$, with conjugate momenta
\begin{equation}
[ X_{ij}, \Pi_{kl} ] = i \delta_{il} \delta_{jk}\ ,\quad [\phi_i, \pi_j] = i \delta_{ij}\ .
\end{equation}
The Hamiltonian is
\begin{equation}
H = \frac{1}{2} {\rm Tr}(\Pi^2) + \frac{m^2}{2} {\rm Tr}(X^2) +  \pi^\dagger (1+gX/M) \pi
+ M^2 \phi^\dagger(1+gX/M) \phi\ .
\end{equation}
In terms of the lowering operators for the fundamental and antifundamental,
\begin{equation}
a_i = \frac{ \pi_i^\dagger  - i M\phi_i}{\sqrt{2M}}\ ,\quad
\bar a_i = \frac{ \pi_i  - i M\phi_i^\dagger}{\sqrt{2M}}\ ,
\end{equation}
this is (dropping a constant)
\begin{equation}
H = \frac{1}{2} {\rm Tr}(\Pi^2) + \frac{m^2}{2} {\rm Tr}(X^2) + M(a^\dagger a + \bar a^\dagger \bar a) + g (a^\dagger X a + \bar a^\dagger X^T \bar a)\ . \label{ham}
\end{equation}
We do not impose the singlet constraint for now, and so there is no Hawking-Page transition.  In Sec.~3.4 we will extend this model to include these features.

There are several motivations that lead us to this model.  The model of Ref.~\cite{Festuccia:2006sa} is based on iteration of the basic graphical unit shown in Fig.~1.
\begin{figure}
\vskip -.4in
\center \includegraphics[width=20pc]{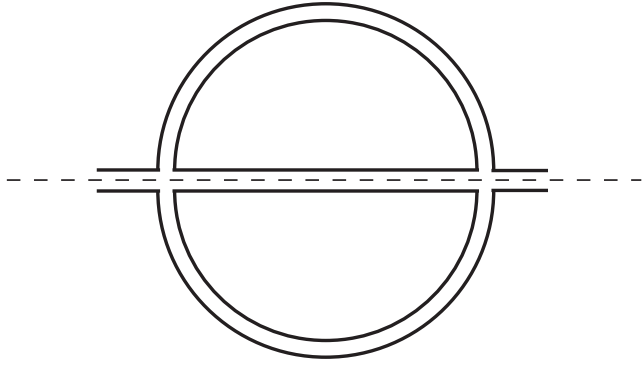}
\caption[]{The basic graphical unit studied in Ref.~\cite{Festuccia:2006sa}.  Iteration of this leads to breakdown of perturbation theory at long times.  Our model iterates a basic unit which is just one side of this, above the dashed line. }
\end{figure}
In our case, the basic process is just the upper half of this graph, above the dashed line: a vector emitting and reabsorbing an adjoint.  The doubling of the graph in Ref.~\cite{Festuccia:2006sa} plays no essential role, so our model should have similar properties.  Moreover, we will see that for us the iteration represents the full planar approximation, whereas in Ref.~\cite{Festuccia:2006sa} it is just a partial summation, while the full planar summation would be much harder.  For the purpose of systematic study of the breakdown of the $1/N$ approximation, it is useful that one can sum the full planar amplitude.

In fact, a very similar model has already been studied in Refs.~\cite{Iizuka:2001cw,Iizuka:2002wa} as an approximation to the quantum mechanics of D0-branes.  That model is more elaborate, in that the fields carry additional indices, and there are fermions, with supersymmetry.  However, the basic graphical structure is the same; in particular the self-interaction of the adjoints is replaced with a quadratic potential, via a mean field approximation.
In Refs.~\cite{Iizuka:2001cw,Iizuka:2002wa}, the adjoints form a D0-brane black hole, and the fundamental is a string stretched from a probe D0-brane to the black hole.  We will be studying the correlator in the one-fundamental sector.  Essentially, we are looking at waves traveling on the stretched string, particularly as they fall into the black hole.  (Ref.~\cite{Lawrence:1993sg} considered a similar situation.)

This system~(\ref{ham}) has no ground state, because the highest term is cubic in the fields.  However, the Hamiltonian commutes with the number operators $N_\phi = a^\dagger a$ and $N_{\bar\phi} = \bar a^\dagger \bar a$, and in each $(N_\phi, N_{\bar\phi})$ sector there is a ground state.  Defining
\begin{equation}
H' = H + c(N_\phi + N_{\bar\phi})(N_\phi + N_{\bar\phi} - 1)\ , \label{hprime}
\end{equation}
the eigenstates of $H'$ are the same as those of $H$, and for sufficiently large $c$ and $M$ the ground state will be in the sector $N_\phi =  N_{\bar\phi} = 0$.  We assume this henceforth.

In fact, for our purposes we need study only the behavior of a single particle in the fundamental represention, in interaction with the matrix heat bath.  To isolate this we will take the splitting $M$ for the fundamental oscillator to be large compared to all other scales, in particular the temperature, and we focus on the observable
\begin{equation}
e^{iM(t-t')} \left\langle {\rm T}\, a_i(t) a_j^\dagger(t') \right\rangle_T
\equiv \delta_{ij} G(T,t-t')
\ . \label{correlator}
\end{equation}
Note that $t$ and $t'$ are Lorentzian, and that in the present section we are interested in temperature $T = 0$.  By construction, the stabilizing term~(\ref{hprime}) vanishes in the relevant sectors $N_\phi = 0,1$, $N_{\bar\phi} = 0$, and so we can calculate with the original $H$~(\ref{ham}).
Including the phase factor in the correlator, the dependence on $M$ drops out in the large-$M$ limit.  Thus there are essentially two parameters at zero temperature, $m$, and $g$ with units of $m^{3/2}$.
In terms of the analogous brane system, taking $M$ to be large means that the probe brane is far from the black hole.

\subsection{The Schwinger-Dyson equation}

At zero temperature, the sum of all planar contributions to the correlator~(\ref{correlator}) is given by the Schwinger-Dyson equation
\begin{equation}
\tilde G(\omega) = \tilde G_0(\omega) - \lambda \tilde G_0(\omega) \tilde G(\omega)
\int_{-\infty}^\infty \frac{d\omega'}{2\pi} \tilde G(\omega') \tilde K_0(\omega - \omega')\ ,
\end{equation}
where $\lambda = g^2 N$ and
\begin{equation}
\tilde G_0(\omega) =  \frac{i}{\omega + i\epsilon}\ ,\quad \tilde K_0(\omega) =  
 \frac{i}{\omega^2 - m^2 + i\epsilon}\ .
\end{equation}
This is shown graphically in Fig.~2.
\begin{figure}
\center \includegraphics[width=30pc]{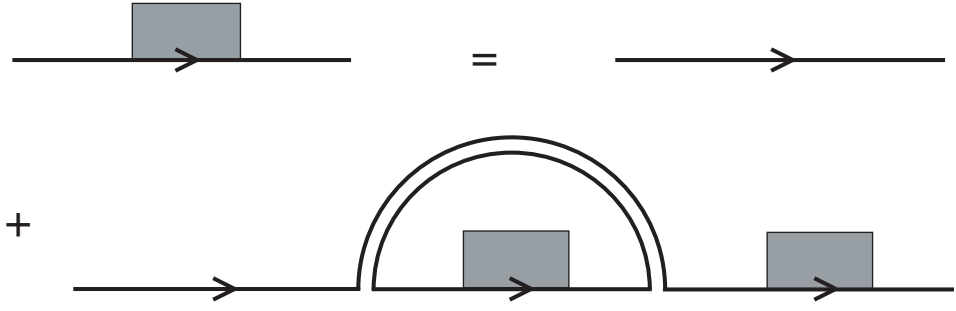}
\caption[]{Schwinger-Dyson equation for planar contributions to $\tilde G(\omega)$ (propagator with shaded rectangle) in terms of $\tilde G_0(\omega)$ and $\tilde K_0(\omega)$. 
} 
\end{figure}
This has the same form as in 2-D QCD~\cite{'t Hooft:1974hx} because the index structure of the interaction is the same.
The integral can be carried out.  Because $a$ annihilates the vacuum, $G(t)$ vanishes for $t < 0$ and so $\tilde G(\omega)$ is nonsingular in the upper half-plane.  Also, because the coupling $g$ has mass dimension $3/2$ we can assume that $\tilde G$ has its free behavior $i/\omega$ at high frequency.  We can then close the contour in the upper half-plane and evaluate the residue at $\omega' = \omega - m + i\epsilon$ to obtain
\begin{equation}
\tilde G(\omega)  = \frac{i}{\omega } 
\left( 1 - \frac{\lambda}{2m} \tilde G(\omega) \tilde G(\omega - m)
\right)\ . \label{sde2}
\end{equation}
We omit the $i\epsilon$ from this equation, with the understanding that the correlator is to be evaluated infinitesimally above the real axis.

In Sec.~2.3 we will solve this in closed form, but first let us study it using a variety of analytic and numerical approaches.
To get our bearings, let us consider first the limit $m \to 0$ with $2\lambda/m \equiv \nu^2$ fixed.  Eq.~(\ref{sde2}) becomes the algebraic equation
\begin{equation}
{\nu^2} \tilde G^2(\omega)  - 4 i \omega \tilde G(\omega)  - 4 = 0\ , \label{alg}
\end{equation}
and so
\begin{eqnarray}
\tilde G(\omega) &=& \frac{2i}{\nu^2} \left(
\omega - \sqrt{\omega^2 - \nu^2} \right)
\nonumber\\
&=& \frac{2i}{\omega + \sqrt{\omega^2 - \nu^2}}\ . \label{sol1}
\end{eqnarray}
On the physical sheet, the square root approaches $\omega$ at long distance.
The $\omega = 0$ pole has been broadened into a branch cut of width $2\nu$.  

This has a simple interpretation.  We are taking the harmonic oscillator frequency $m$ to 0, so the matrix $X$ is essentially a static variable with an eigenvalue distribution given by the Wigner semi-circle law.  The propagator~(\ref{sol1}) is just the free propagator with mass $\mu = gX$, averaged over a semi-circle eigenvalue distribution for $X$ of width $\sqrt{2N/m}$.  
The branch cuts in $\omega$ translate into the asymptotic power law decay $e^{ \pm i \nu t} t^{-3/2}$.  However, this power law decay is unrelated to the quasinormal behavior that we seek: it originates from the noncompactness of $X$ due to the vanishing of the potential at $m = 0$, rather than the large-$N$ limit.  

Writing Eq.~(\ref{sde2}) as a recursion relation
\begin{equation}
\frac{1}{\tilde G(\omega)} =  \frac{\nu^2}{4} \tilde G(\omega - m)  -{i}{\omega}
\ . \label{rec1}
\end{equation}
gives an efficient way to calculate numerically, beginning with the asymptotic behavior~$\tilde G(\omega) \sim i /(\omega + i\epsilon)$.  In order this procedure to work, we must require that the recursion be stable at large $|\omega|$ --- otherwise, subasymptotic terms could grow to become significant.  Suppose that there is a solution $\tilde G_*(\omega)$, and we consider a perturbation $\tilde G_*(\omega) + \gamma(\omega)$.  Then
\begin{equation}
\gamma(\omega) =  -\frac{\nu^2}{4}  \tilde G_*^2(\omega) \gamma(\omega - m) 
\ . \label{rec2}
\end{equation}
The recursion is stable towards increasing $\omega$ when $|{\nu^2}  \tilde G_*^2(\omega)| \leq 4$, and stable towards decreasing $\omega$ when the inequality is reversed.
Since $\tilde G(\omega) = O(\omega^{-1})$ at large $|\omega|$, the recursion is stable there in the direction of increasing $\omega$.  Given perfect numerical precision, we could start at very large negative $\omega$ and use the convergence to bring us very close to a solution.  Even if the recursion became unstable for some intermediate range, we could make the initial error as small as desired.  In practice, a highly unstable recursion would lead to a numerical errors.  This is an potentially an issue at small $m$, where the recursion requires many steps.  In fact, using the $m=0$ solution~(\ref{sol1}) one finds that the recursion to the right is no worse than neutrally stable in this limit.

Before performing the numerics, we can anticipate that the branch cut found at  $m=0$ must break up into poles.  Since there is no branch cut at large $\omega$, none can appear through the recursion.  However, whenever the right-hand side of Eq.~(\ref{rec1}) vanishes, there will be a pole in $\tilde G(\omega)$.  Generically this will happen at isolated points on the real axis, because the right-hand side is purely imaginary there.  The numerical integration verifies this picture: the branch cut breaks up into poles as $m$ is turned on, and the poles move further apart as $m$ is increased with $\nu$ fixed.  Zero temperature results for the real part of $\tilde G(\omega)$ are shown in Fig.~3 for $\nu = 1$, $m=0.05$ and in Fig.~4a for $\nu = 1$, $m=0.80$.
\begin{figure}
\center \includegraphics[width=35pc]{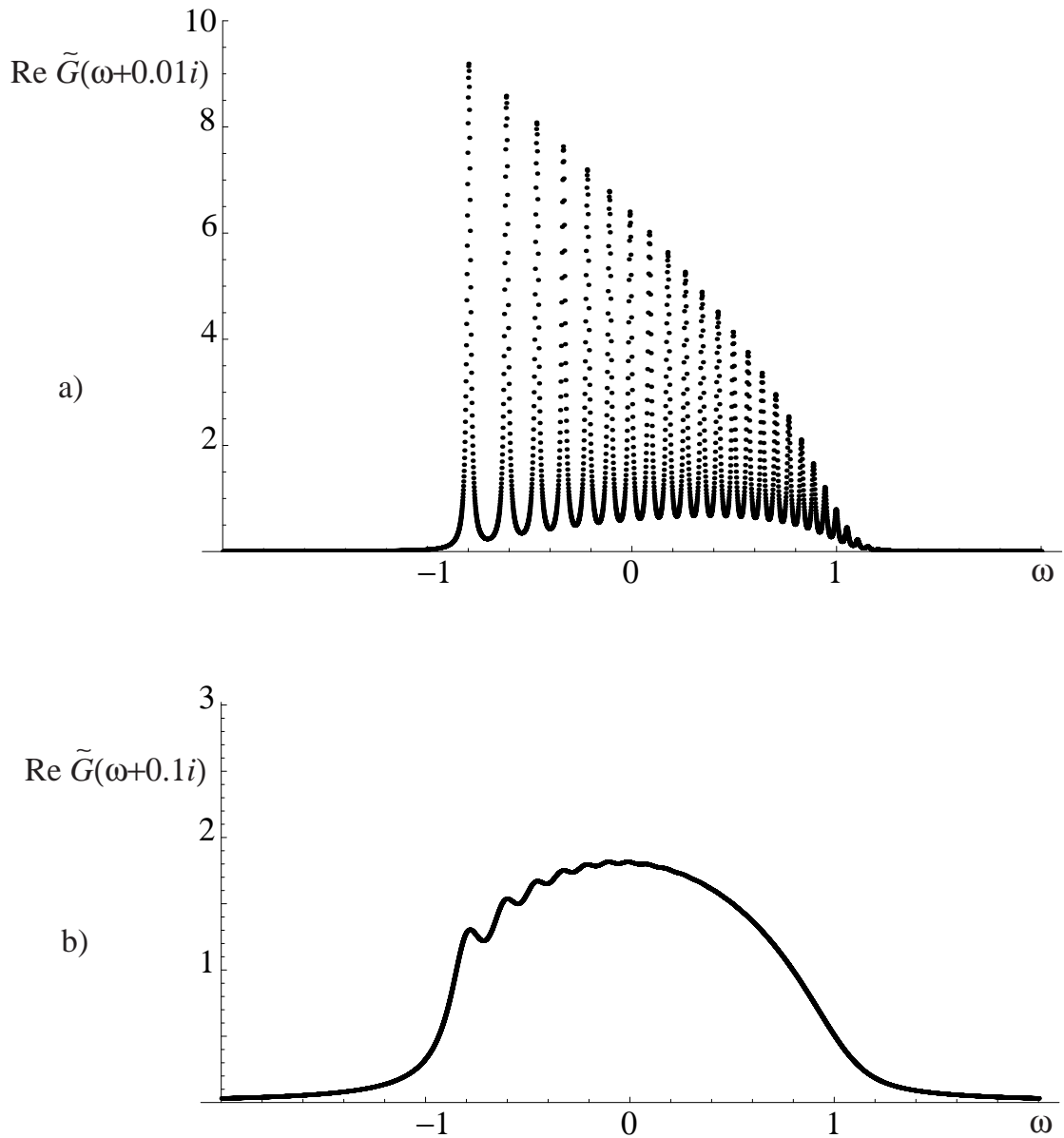}
\caption[]{a) The real part of $\tilde G(\omega)$ for $\nu = 1$, $m=0.05$, evaluated 0.01 units above the real axis to give the delta functions finite width.  b) The same function evaluated 0.1 units above the real axis: the poles merge into an approximate semicircle distribution.} 
\end{figure}

There is a temptation at small $m$ to approximate the recursion~(\ref{rec1}) by a differential equation, but this does not seem to be useful: the differential equation is less stable than the recursion relation.  They agree on smooth configurations, but have different $e^{O(\omega/m)}$ instabilities.

\subsection{Canonical calculation}

The intermediate states in Fig.~2 consist of a single $\phi$ excitation plus any number of $X$ excitations, with the indices contracted:
\begin{equation}
|j,r\rangle =  N^{-r/2} a^\dagger_i {(A^{\dagger r})_{ij}} |v\rangle\ , \quad r \geq 0\ ,\label{inter}
\end{equation}
where $ |v\rangle$ is the free ground state and $A^\dagger_{ij} = ( \Pi_{ij}  + i mX_{ij})/\sqrt{2m}$ is the raising operator.  The states $|j,r\rangle$ are orthonormal at large $N$.  Then
\begin{equation}
(H - M)|j,r\rangle =  mr |j,r\rangle + \frac{i\nu}{2} |j,r+1\rangle - \frac{i\nu}{2} |j, r-1\rangle
-  \frac{i\nu}{2} \sum_{l=1}^{r-1} N^{-1-l/2} (A^{\dagger l})_{kk}  |j,r-l-1\rangle\ .
\end{equation}
The last term has norm of order $N^{-1}$ and so drops out in the planar limit.  The noncompactness in $r$ suggests that this model may show quasinormal behavior at finite temperature.

Defining 
\begin{equation}
|j,\psi\rangle = \sum_{r=0}^\infty \psi_r |j,r\rangle\ ,
\end{equation}
we have the eigenvalue condition
\begin{equation}
(\omega - mr) \psi_r = \frac{i\nu}{2} (\psi_{r-1} - \psi_{r+1})\ , \quad \psi_{-1} \equiv 0\ .
\end{equation}
This is essentially the Bessel recursion relation.  Specifically,
\begin{equation}
\psi_r = i^{-r} J_{r-\omega/m}(\nu/m)\ . \label{wavefun}
\end{equation}
Here $i^{-r} N_{r-\omega/m}(\nu/m)$ would satisfy the same recursion relation, but only the solution~(\ref{wavefun}) is normalizable.
The eigenvalue condition $J_{-1-\omega/m}(\nu/m) = 0$ determines the poles in the correlator.  With this clue, we can find the closed-form solution to the Schwinger-Dyson equation~(\ref{sde2}),
\begin{equation}
\tilde G(\omega) = \frac{2i}{\nu}\frac{ J_{-\omega/m}(\nu/m)}{J_{-1-\omega/m}(\nu/m)}\ .
\end{equation}

\section{Finite temperature}

\subsection{The Schwinger-Dyson equation}

Now consider the system at nonzero temperature.  To study real-time thermal correlators one generally needs the doubled integration contour of the Schwinger-Keldysh formalism~\cite{Schwinger:1960qe,Keldysh:1964ud,Niemi:1983nf}.  However, our situation simplifies.  First, we are assuming that $M$ is large compared to the temperature, so there are no $\phi$ excitations in thermal equilibrium.  Second, the $X$ fields have no self-interaction, so the thermal ensemble is free.\footnote{Note that in any event the backreaction of the fundementals on the adjoints is suppressed in $N$.}  In this case the real-time formalism reduces to replacing the $X$ propagator with the free thermal propagator,
\begin{equation}
\tilde K_0(T,\omega) = \frac{i}{1 - e^{- m/T}} \left( \frac{1}{\omega^2 - m^2 + i\epsilon}
- \frac{e^{- m/T}}{\omega^2 - m^2 - i\epsilon} \right) \ . \label{ktherm}
\end{equation}
The Schwinger-Dyson equation is changed only by the use of this propagator,
\begin{equation}
\tilde G(T,\omega) = \tilde G_0(\omega) - \lambda \tilde G_0(\omega) \tilde G(T,\omega)
\int_{-\infty}^\infty  \frac{d\omega'}{2\pi} \tilde G(T,\omega') \tilde K_0(T,\omega - \omega')\ .
\end{equation}
The time-ordered correlator $G(T,t)$ still vanishes at $t < 0$ (because $ M \gg T$) and so we can again close the contour in the upper half-plane and pick up only the poles of $\tilde K_0(T,\omega) $.  This gives
\begin{equation}
\tilde G(T,\omega)  = \frac{i}{\omega  } 
\left\{ 1 - \frac{\nu^2}{4(1 - e^{- m/T})} \tilde G(T,\omega) \left[\tilde G(T,\omega - m) + e^{- m/T} \tilde G(T,\omega + m) \right]
\right\}\ , \label{sde3}
\end{equation}
or
\begin{equation}
\tilde G(T,\omega - m)  - \frac{4}{\nu_T^2}\frac{1}{ \tilde G(T,\omega)} + e^{- m/T} \tilde G(T,\omega + m) 
= \frac{4 i \omega}{\nu_T^2}  \label{sde4}
\end{equation}
with $\nu_T^2 = \nu^2/(1- e^{- m/T})$.

Although this is similar to the zero-temperature Schwinger-Dyson equation, the behavior of its solutions is very different.  At zero temperature the singularities of $\tilde G(\omega)$ are poles on the real axis.  At nonzero temperature such poles are impossible.  

To see this, note first the spectral representation,
\begin{equation}
G(T,t-t') =  {\cal N} {\rm Tr} \Bigl(e^{- H/T} a_1(t) a_1^\dagger(t') \Bigr)=
  {\cal N} \sum_{A,B} |\langle A | a_1 | B \rangle |^2 \theta(t-t') e^{- E_A/T - i (t-t')(E_B - E_A)}\ ,
\end{equation}
where ${\cal N}^{-1} =  {\rm Tr}\, e^{- H/T}$.  The Fourier transform is
\begin{eqnarray}
\tilde G(T,\omega) &=&  i   {\cal N}\sum_{A,B} \frac{ e^{- E_A/T} |\langle A | a_1 | B \rangle |^2}{\omega - E_B + E_A } 
\nonumber\\
 &=& i \int_{-\infty}^\infty \frac{d\mu\, \pho(\mu)}{\omega - \mu }\ . \label{spec}
\end{eqnarray}
We have introduced the nonnegative spectral density
\begin{equation}
\pho(\mu) =   {\cal N} \sum_{A,B} { e^{- E_A/T} |\langle A | a_1 | B \rangle |^2}\delta(\mu - E_B + E_A ) \ .
\end{equation}
Now, suppose that there is a pole in $\tilde G(T,\omega)$ at some $\omega_0$.  Then the first term on the LHS of Eq.~(\ref{sde4}) has a pole at $\omega = \omega_0 + m$.  This must be cancelled either by a pole in the last term with a residue of the opposite sign, or by a pole in the second term from a zero of $\tilde G(T,\omega)$ at $\omega = \omega_0 + m$.  The spectral representation forbids a negative residue, so the zero must exist.  Similarly we can conclude that there is a zero at $\omega_0 - m$.  It then follows that all three terms on the LHS vanish at $\omega = \omega_0$, which is a contradiction except possibly for a pole at $\omega_0 = 0$ (which is indeed present in the free theory).

The absence of poles, and so of delta-functions in $\pho(\mu)$, immediately implies that the correlator goes to zero asymptotically in time: this model has the planar behavior that we seek.  If the zero temperature poles separate into branch cuts on the real axis there will be power law falloff; if the singularities drop below the real axis onto the second sheet the decay will be exponential.  We find numerically that at small nonzero temperature the poles widen into branch cuts, while at higher temperature the cuts merge and the singularities drop below the axis.  We will present these results in Sec.~3.2, after some further analytic discussion.

The recursion relation again becomes algebraic in the limit $m \to 0$, now with $\nu_T^2$ fixed,
\begin{equation}
(1+e^{- m/T}){\nu_T^2} \tilde G^2(T,\omega)  - 4 i \omega \tilde G(T,\omega)  - 4 = 0\ . \label{alg2}
\end{equation}
The solution is 
\begin{equation}
\tilde G(T ,\omega) = \frac{2i}{(1+e^{- m/T})\nu_T^2} \left(
\omega - \sqrt{\omega^2 - (1+e^{- m/T}) \nu_T^2} \right) \ . \label{m0t}
\end{equation}
The logic is the same as at zero temperature, with the eigenvalue distribution thermally broadened. 

The three term recursion relation~(\ref{sde4}) is unstable in both directions, so its solution is less constrained than at zero temperature.  We can derive some useful results from the real part ${\rm Re}\,\tilde G(T,\omega) =  \pi \pho(\omega)$.  Then
\begin{equation}
\pho(\omega - m)  - \frac{4}{\nu_T^2 | \tilde G(T,\omega) |^2} {\pho(\omega)}{} + e^{- m/T} \pho(\omega + m) = 0 \ .
\label{real} 
\end{equation}
First, if the spectral density has support in any segment of the real axis then it has support in every segment translated by a multiple of $m$.  Thus, when the poles spread into branch cuts, each cut is accompanied by an unbounded series of additional cuts.  This does not contradict the known asymptotic behavior, because the magnitude of $\pho(\omega)$ goes to zero at large $\omega$.  The form of the falloff follows from the fact that coefficient of the middle term in Eq.~(\ref{real}) becomes large asympotically, $4 \omega^2/\nu_T^2$.  The recursion relation is then dominated by two terms,
\begin{eqnarray}
\pho(\omega-m)/\pho(\omega) &\cong& e^{- m/T} \nu_T^2/4\omega^2 \ ,\quad 
\omega \to -\infty\ ,\nonumber\\
\pho(\omega+m)/\pho(\omega) &\cong& \nu_T^2/4\omega^2 \ ,\quad 
\omega \to +\infty\ ,\label{ttr}
\end{eqnarray} 
so the spectral density behaves asymptotically as $|\omega|^{-O(|\omega|)}$.

\subsection{Numerical results}

The instability of the recursion relation makes numerical calculation challenging.  One approach would be to fine tune the initial condition.  However, we have had more success by a different approach, solving the stable zero temperature recursion relation and then solving the coupled differential equations obtained by differentiating the recursion relation~(\ref{sde4}) with respect to $T$ at fixed $\nu_T$.  Figs. 4 show $F(\omega)$ for a series of temperatures from zero to infinity, with $\nu_T = 1$ and $m = 0.8$.
\begin{figure}
\vskip -.5in
\center\includegraphics[width=19pc]{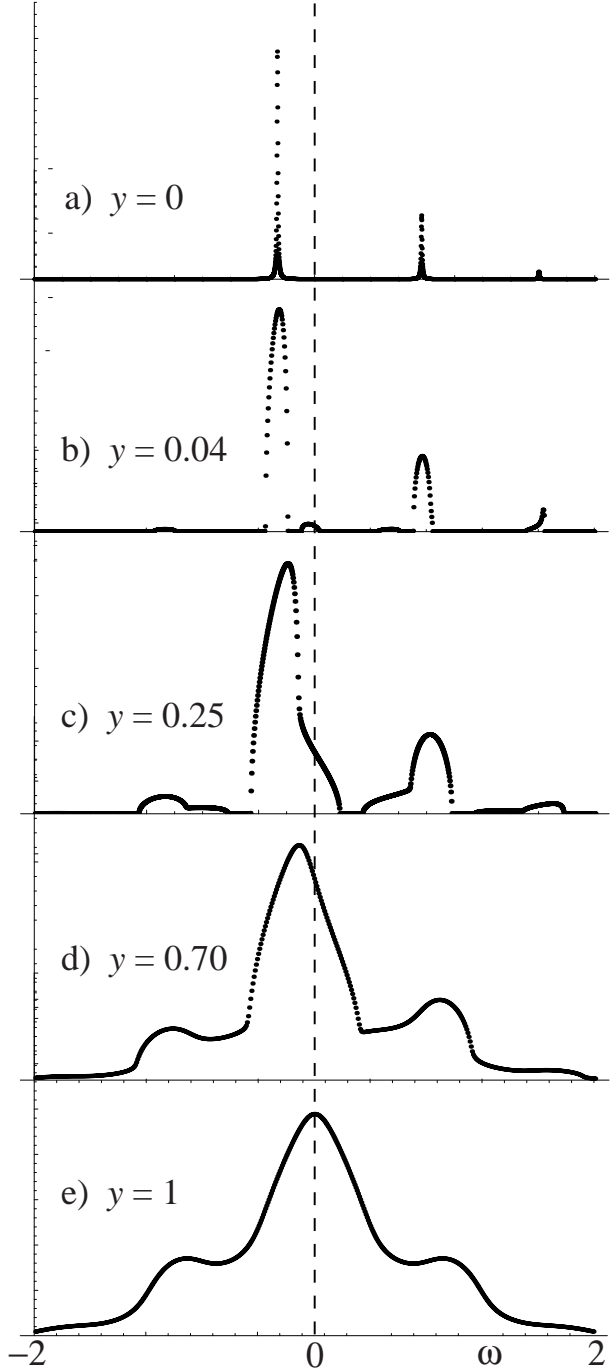}
\caption[]{The real part of $\tilde G(\omega)$ for $\nu_T = 1$, $m=0.80$, and various values of $y = e^{- m/T}$.  The vertical axis is rescaled at each temperature for best visibility (the actual area under the curve is $\pi$ at all temperatures), and at zero temperature $\omega$ is taken slightly above the real axis for the same reason.} 
\end{figure}

At zero temperature (a) this is a set of delta functions, meaning poles in $\tilde G(\omega)$.  At low temperature (b) the poles become short branch cuts, and additional cuts appear, shifted by multiples of $m$.  As the temperature is increased (c) the cuts lengthen and begin to merge, but there are still gaps.  At higher temperatures (d) the cuts have completely merged and $F(\omega)$ is everywhere positive and smooth: the singularities have moved onto the second sheet.  The recursion relation and its solution have a sensible infinite temperature limit (e).  In the infinite temperature limit the equation is symmetric under $\tilde G(\omega) \to \tilde G(-\omega^*)^*$, and this symmetry provides a check on the numerical approach. We have verified numerically that the high-temperature solution is smooth out to $|\omega| =4$, and the asymptotics~(\ref{ttr}) imply that nonanalyticity cannot arise at large $\omega$ if not present at smaller values.

\begin{figure}
\vskip -1in
\center\includegraphics[width=22pc]{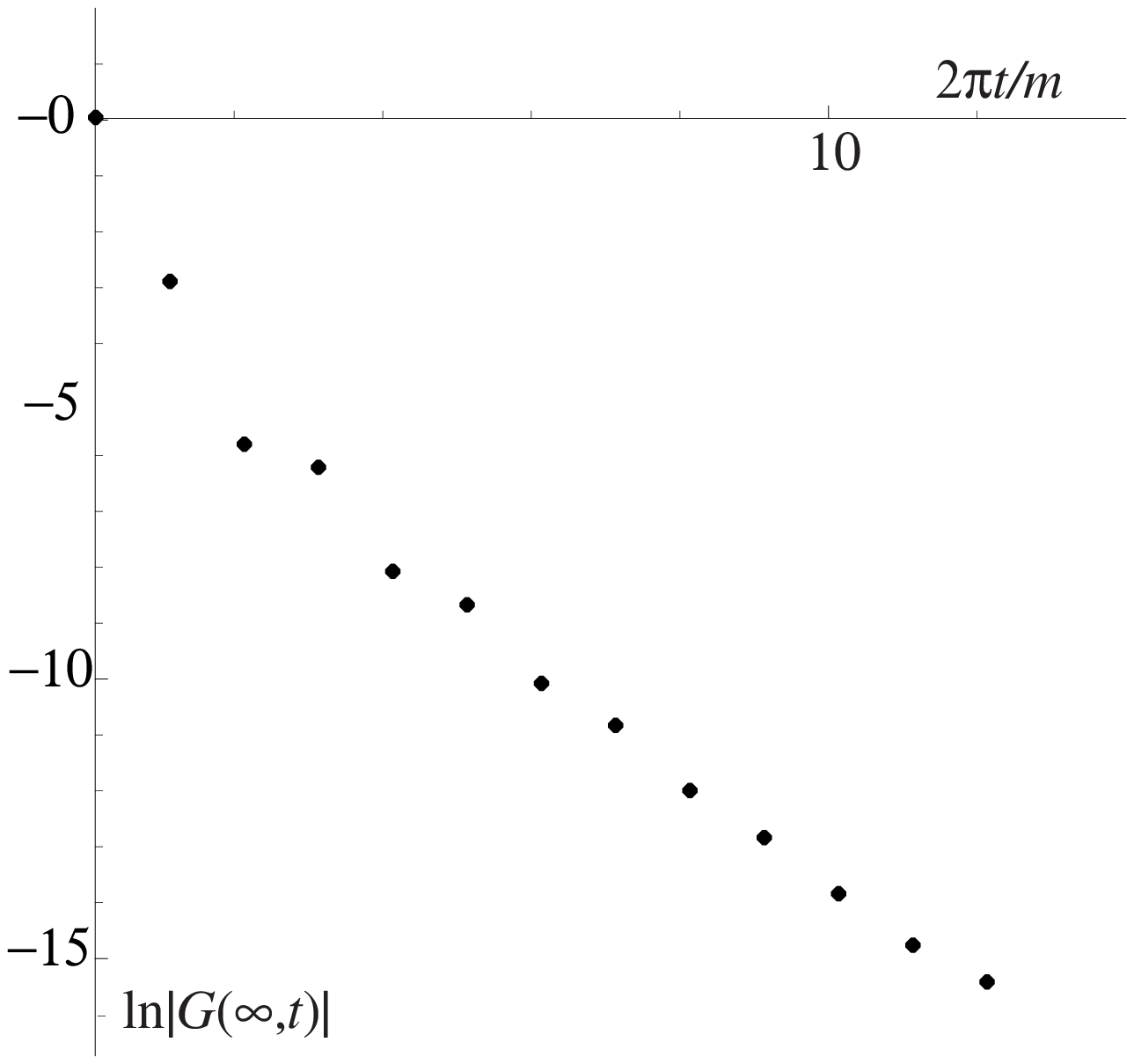}
\caption[]{The logarithm of the real time infinite temperature correlator, $\ln |G(t)|$, for $\nu_T = 1$, $m =0.8$, and $T= \infty$.  For clarity, time differences of $2\pi/m$ are displayed: on shorter intervals the correlator shows strong oscillations due to interference of singularities spaced by multiples of $m$.} 
\end{figure}
The Fourier transform of Fig.~4e is shown in Fig.~5.  Two exponentials are evident. The amplitude and decay rate of the first few points matches the central peak in Fig.~4e.  The long-lived exponential arises from the closest pole to the real axis, which has a residue smaller by a factor of order $10^{-2}$.  This appears to be associated with the last branch cuts to merge, and with the kink near $\omega = 0.3$ in Fig.~4d (examination with greater resolution verifies that the function in Fig.~4d is smooth at this point).

\subsection{Some approximations}

Thus, our simple model has the behavior that we seek.  It would be good to have an approximate analytic treatment of the quasinormal behavior.  This is difficult because the model has three energy scales, $m$, $\nu_T$, and $T$, and the quasinormal behavior disappears if any are set to zero.\footnote{There is power law decay at $m=0$, but as we have noted this is not associated with large $N$.}  Note that in Fig.~4 we have taken $\nu_T = 1$, $m = 0.8$, near the middle of parameter space, because the numerics are cleanest there.

We can develop an approximate solution when any of the parameters is small; here we note a few features of two such approximations.

\subsubsection{Small $\lambda$ (small $\nu_T$)}

The ordinary perturbation theory in $\lambda$ is singular here at long times~\cite{Festuccia:2006sa}.  In our model, this shows up as the fact that each additional order of perturbation theory brings additional poles in $\tilde G(T,\omega)$, even within the 1PI part.  However, we can obtain an improved approximation from what we know of the solution.  We expect that the pole at $\omega = 0$ will turn into a short branch cut, and that additional branch cuts will develop translated by multiples of $m$.  From the structure of the recursion relation, the spectral weight in the branch cut at $\omega = mk$ will is of order $g^{-2|k|}$.  Again, $\nu_T^2 = 2\lambda/m(1-e^{- m/T})$.  To see the branch cut we focus on $\omega \ll m$, for which the recursion relation implies
\begin{eqnarray}
\tilde G(T,\omega - m)  - \frac{4}{\nu_T^2}\frac{1}{ \tilde G(T,\omega)} + e^{- m/T} \tilde G(T,\omega + m) 
&=& \frac{4 i \omega}{\nu_T^2} \ ,\nonumber\\
\tilde G(T,\omega)  - \frac{4}{\nu_T^2}\frac{1}{ \tilde G(T,\omega+m)} 
&=& \frac{4 i m}{\nu_T^2} \ ,\nonumber\\
-\frac{4}{\nu_T^2}\frac{1}{ \tilde G(T,\omega-m)} + e^{- m/T} \tilde G(T,\omega) 
&=& -\frac{4 i m}{\nu_T^2} \ . \label{0pm}
\end{eqnarray}
In the second and third lines we have have used $\omega \ll m$ on the RHS, and have dropped one term on the LHS because $G(T,\omega \pm 2m) \ll G(T,\omega)$.  Using these we can eliminate  $G(T,\omega \pm m)$ to obtain the cubic equation
\begin{equation}
i\omega y \tilde G(T,\omega)^3  + (-1+y-y^2 -\omega\tau + y\omega\tau) \tilde G(T,\omega)^2 + i\tau(2 - 2y + \omega\tau) \tilde G(T,\omega ) +\tau^2 = 0\ ,
\end{equation}
where $y = e^{- m/T}$ and $\tau = 4m/\nu_T^2$.  For simplicity consider the limit $y \to 1$, where
\begin{equation}
i\omega \tilde G(T,\omega)^3  - \tilde G(T,\omega)^2 + i \omega\tau^2 \tilde G(T,\omega ) +\tau^2 = 0\ .
\end{equation}
At $\omega = 0$, $\tilde G(T,\omega) = \tau$ is purely real.  However, the discriminant for this equation vanishes at $\omega^2 = (11 + 5^{3/2})/2\tau^2$, beyond which the solution is purely imaginary. This determines the branch cut width $O(\tau^{-1}) = O(\lambda/m^2 [1 - y])$.  

These results have been confirmed numerically.  Note that the spectral density implied by the cubic equation is more complicated than the semicircle law that might have been expected for a short cut.\footnote{The discriminant also vanishes at  $\omega^2 = (11 - 5^{3/2})/2\tau^2$, so there is another branch cut not far below the real axis.} 
The cuts at $\omega \pm m$ are also determined by the relations~(\ref{0pm}) and are smaller by $O(\lambda)$.  The further cuts are given by the two-term recursion~(\ref{ttr}) and are down by $\lambda^{|k|}$ as expected.  One can systematically carry this small-$\lambda$ approximation to higher order and no singularities arise, so we have found the quantitative small-$\lambda$ behavior.  The branch points imply power law decay, so there is again a contradiction with the finite-$N$ behavior.

\subsubsection{Small $m$}

At small $m$ the recursion relation goes over naively to the differential equation
\begin{equation}
m (1-e^{-m/T}) \tilde G(T,\omega)\tilde G(T,\omega)' =   (1 + e^{- m/T})  \tilde G(T,\omega)^2 - \frac{4 i \omega}{\nu_T^2}  \tilde G(T,\omega) - \frac{4}{\nu_T^2}\ ,
\end{equation}
where the prime denotes an $\omega$-derivative.
We are holding $m/T$ fixed.  If $T$ were held fixed the LHS would be of order $m^2$ and we would need to keep also a second derivative term, but the conclusions would be similar.  The derivative term is a perturbation, but a singular one~\cite{BO}: it can become large in regions of large gradient.  Again it is useful to study the stability near a solution $\tilde G_*$, expanding $\tilde G = \tilde G_* + \gamma$:
\begin{equation}
m (1-e^{-m/T}) \tilde G_* \gamma' =   - m (1-e^{-m/T}) \tilde G_*' \gamma +  2(1 + e^{- m/T})  \tilde G_* \gamma - \frac{4 i \omega}{\nu_T^2}  \gamma\ .
\end{equation}
Assuming that the solution $G_*$ is close to the algebraic solution~(\ref{m0t}), one finds that the solution is stable toward increasing $\omega$ for $\omega < - \hat\nu$ and for $\omega >  \hat \nu$, where
$\pm\hat\nu = \pm \nu_T \sqrt{1+e^{-m/T}}$ are the branch points of the $m=0$ solution, and stable toward decreasing $\omega$ for $-\hat\nu < \omega < \hat\nu$.

Thus, starting from the known asymptotic behavior at large negative $\omega$, the differential equation has a unique solution which is close to the $m=0$ solution for $\omega < -\hat\nu$ and then blows up rapidly, as $e^{\omega \hat\nu}$.  Similarly by starting at  $\omega = \hat\nu$ and integrating in both directions one obtains a solution that is close to the $m=0$ solution for  $\omega > -\hat\nu$ and blows up beyond this point.  These solutions clearly do not match onto one another as a single solution to the differential equation.  The point is that near the branch point $\omega = -\hat\nu$  the gradients become of order $1/m$ and we must use the original discrete form of the equation.  The full solution requires matching this `boundary layer' solution with the smooth outer solutions that we have found.  The details are an interesting direction for future work.  In particular, the singularities will be closest to the real axis in the boundary layer, where the variation is the most rapid, and so this part of the solution dominates the long-time behavior.

\subsection{The Hawking-Page transition}

In the model thus far, we have not imposed the singlet constraint that generally is present in quantum mechanical realizations of gauge/gravity duality.  In consequence, there is no phase transition as the temperature is varied.  These features can be restored without altering the earlier results, by introducing some decoupled sectors.  First, to form a singlet we need an antifundamental excitation, which we can obtain from an additional decoupled oscillator.  It would be more natural to excite an antifundamental excitation of the original $\phi$ oscillator, but then we would also need to include ladder graphs as in Ref.~\cite{'t Hooft:1974hx}; this may be an interesting direction for future work.

The thermal phase transition with large-$N$ oscillators has been studied in Ref.~\cite{Gao:1998ww} and in more detail in Ref.~\cite{Aharony:2003sx}.  We will review this very briefly.  The singlet constraint is enforced by integrating over a Wilson-Polyakov line $U$ in the Euclidean time direction.  At fixed diagonal $U$ the adjoint thermal propagator is
\begin{eqnarray}
&&\int_{-\infty}^\infty dt\, e^{i\omega t} \langle {\rm T}\, X_{ij}(t) X_{kl}(0) \rangle
= \delta_{il}\delta_{jk} \Biggl\{
\frac{i}{\omega^2 - m^2 + i\epsilon}
\nonumber\\
&&\qquad\qquad
+ \frac{\pi}{m} \sum_{n=1}^\infty \left[
(e^{- m/T}U_{ii}U_{jj}^{-1})^n \delta(\omega+m) + (e^{- m/T}U_{jj}U_{ii}^{-1})^n \delta(\omega-m) \right] \Biggr\}\ .  \label{uprop}
\end{eqnarray}
The sum represents paths that wind around the Euclidean time direction $\pm n$ times, each winding picking up a phase from the Wilson-Polyakov line.
All adjoint propagators in a graph are evaluated with the same $U$, which is then integrated over.
The $\phi$ propagator is unaffected because $M$ is large: paths that wind around the Euclidean time direction do not contribute.

We extend the previous model by the addition of an adjoint oscillator of frequency $m'$.  This is decoupled from the other fields except through the singlet constraint, that is, through its coupling to $U$.  The phase transition occurs when~\cite{Aharony:2003sx}
\begin{equation}
e^{- m/T_c} + e^{- m'/T_c} = 1\ .
\end{equation}
As $m' \to 0$, $T_c \to 0$ and as $m' \to \infty$, $T_c\to \infty$, so we can adjust the transition to take place at any temperature.  

At $T < T_c$, the eigenvalues $U_{ii}$ are uniformly distributed on the unit circle, so that $\sum_i U_{ii}^n = 0$ for any finite $n$.  It then follows that the $U$-dependent terms from Eq.~(\ref{uprop}) drop out in every trace at leading order in $N$, so the correlator reduces to its zero temperature form (see for example Refs.~\cite{Furuuchi:2005eu,Brigante:2005bq})  Thus there is no quasinormal behavior.  

At $T > T_c$ the eigenvalues become nonuniform, and at $T \gg T_c$ they are concentrated around the identity.  For $T \gg m'$ we can replace $U$ with the identity, and the propagator~(\ref{uprop}) reduces to its form~(\ref{ktherm}) without the singlet constraint.  We then recover the quasinormal behavior found earlier.  

\section{The information paradox}

Now let us return to the information paradox, and discuss what we might learn.  Our model does not capture all aspects of the paradox, spacetime locality in particular: like the weakly coupled gauge theory, there is no large bulk dual and no notion of spacetime locality.  At best, the model represents a bulk theory in a large curvature limit.

What the model does preserve is the large-$N$ structure, the existence of dissipative behavior in the planar limit that does not survive at finite $N$.  The $1/N$ expansion in gauge/gravity duality is dual to the loop expansion in quantum gravity.  It is a crucial question, how the preservation of information is manifested in this expansion.  Whatever form the answer takes, there should be some parallel in our model.

For example, there have been many proposals over the years that evidence for information restoration can be seen in the breakdown of the gravitational loop expansion even at low orders; see Ref.~\cite{Giddings:2007ie} for a recent discussion along these lines.  A rather different proposal is that the restoration arises from nonperturbative effects in the gravitational loop expansion, additional saddle points that were omitted in the original argument~\cite{Maldacena:2001kr,Hawking:2005kf}.  These two approaches would correspond respectively to an order-by-order breakdown of the $1/N$ expansion, and to the contribution of $e^{-O(N^2)}$ effects.

The saddle point proposal has been analyzed critically in Refs.~\cite{Birmingham:2001pj,Birmingham:2002ph,Barbon:2003aq,Barbon:2004ce,Kleban:2004rx,Barbon:2005jr}.  We can paraphrase their key argument as follows.  In the planar approximation, in the regime where the singularities lie below the real axis, the spectral density $\pho(\mu) = {\rm Re}\, \tilde G(\mu)/\pi$ has support on the whole real axis.  At finite $N$, this must break up into poles with a typical spacing $e^{-O(N^2)}$.  If we consider an observable corresponding to the convolution of $\tilde G(\mu)$ with some smooth function, the effect is only of order $e^{-O(N^2)}$, and so looks as though it might be captured by a saddle point contribution~\cite{Maldacena:2001kr,Hawking:2005kf}.  However, if we measure $\tilde G(\mu)$ at a precise value of $\mu$, the effect is of {order one} (or larger), which cannot be captured by such a saddle point.  

A Euclidean saddle point gives a contribution to the density matrix, so with two such saddles we have
\begin{equation}
\rho = \frac{\rho_1 + e^{S_1 - S_2} \rho_2}{1 + e^{S_1 - S_2}}\ ,
\end{equation}
with $\rho_1$ and $\rho_2$ normalized density matrices and $S_2 - S_1$ of order $G^{-1} \sim N^2$.
This implies for the correlator
\begin{equation}
F = \frac{F_1 + e^{S_1 - S_2} F_2}{1 + e^{S_1 - S_2}}\ .
\end{equation}
Observing the order one effect requires measurements on a time scale of order $e^{O(N^2)}$, which suggests the possibility of a compensating $N$-dependence.\footnote{For black holes that decay, the time scale is very much shorter.  However, in this case one must observe of order $N^2$ particles, so again there is the possibility of an offsetting factor.}  However, the extra saddle points do not have the necessary enhancement: we need that the total residue of the poles in $F(\mu)$ be 1.  The dominant saddle gives a continuum with total weight $1 - e^{-O(N^2)}$, while the secondary saddles may give poles but with total weight $e^{-O(N^2)}$.

Our discussion suggests a further problem with this idea.  Namely, the difference between the planar and finite-$N$ behavior already shows up in the basic model presented in Sec.~2.1, without the singlet constraint and without the Wilson line variable $U$.  The additional saddle discussed in Refs.~\cite{Maldacena:2001kr,Hawking:2005kf} appears in the $U$-integration.

We should be alert to possible artifacts of our model.  In particular, the fact that we find power law decay at small $g$ for arbitrarily large temperatures is likely connected with the fact that the fields in our adjoint heat bath are free, so states do not mix as completely as they should.  We expect that in the weakly coupled gauge theory (at temperatures above the Hawking-Page transition) there will be exponential decay.  However, the power law decay is already in contradiction with the finite-$N$ behavior, so for our purposes we can regard this as quasinormal behavior.

Because we do not have a bulk spacetime, we cannot separate stringy physics from gravitational field theory.  That is, we capture the gravitational loop expansion but not the $\alpha'$ expansion.  Thus we cannot directly investigate proposals such as those in Refs.~\cite{Susskind:1993aa,Lowe:1995ac}, which relate information recovery to specifically stringy physics.

Our discussion has been entirely in the field theory language.  The dual language would correspond to working with $U(N)$ invariants such as products of $A$ and $A^\dagger$, either traced or in bilinears with $a$ and $a^\dagger$; it would be interesting to develop this further.  In this form, finiteness of $N$ shows up as relations between these invariants; for example, Tr$(A^k)$ can be expressed in terms of lower traces for $k > N$.  One does not expect these relations to be visible in perturbation theory in $1/N$, since for any given $k$ they turn on abruptly at a finite value of $1/N$.  Rather, they seem to imply that the closed string bulk variable are simply not good variables for nonperturbative gravity.

This reduction of the number of independent variables is the `stringy exclusion principle'  \cite{Maldacena:1998bw}.  It has been related to the growth of the size of objects at high energy~\cite{McGreevy:2000cw}.  The information paradox also implies another reduction in the size of the Hilbert space, the `black hole complementarity principle'~\cite{Susskind:1993if}.  For example, in an eternal AdS black hole there is an infinite number of infalling modes at the horizon; however, only a Bekenstein-Hawking entropy's worth can be independent.  The idea that black hole complementarity is a consequence of the stringy exclusion principle is implicit in various places; the formulation of the information paradox in Ref.~\cite{Maldacena:2001kr} makes it particularly clear.

Growth of string states near the horizon was discussed in Refs.~\cite{Susskind:1993ki,Mezhlumian:1994pe}.  Unlike the case studied in Ref.~\cite{McGreevy:2000cw}, where the string blows up into a spherical D3-brane, at the horizon it grows into a randomly walking long string.  In the gauge theory, the former corresponds to the trace of a large power of a single field, while the latter suggests the more generic case of the trace of a random sequence of fields.

D-branes are related to the $(2n_{\rm loop})!$ growth of perturbation theory~\cite{Shenker:1990uf}, versus $n_{\rm loop}!$ in field theory, and to the associated $e^{-1/g_{\rm s}}$ effects (though they cannot be the only such effects~\cite{Silverstein:1996xp}).  Given the connection between the stringy exclusion principle and D-branes, we are led to conjecture that the breakdown of low energy effective field theory in the neighborhood of a black hole is manifested by a $(2n_{\rm loop})!$ growth of perturbation theory, and by $e^{-O(1/\sqrt G)}$ nonperturbative effects, as compared to $e^{-O(1/G)}$ contributions from field theory saddles.  This conjecture is just based on analogy; we do not have any specific scenario for how these large high order amplitudes appear, and for how the $e^{-O(1/\sqrt G)}$ effects become order one in certain observables.

Lastly, we note an interesting recent paper~\cite{Hayden:2007cs}
 on the time scale for black hole information return.  The arguments of this paper rest on certain assumptions about the thermalization process.  It may be possible to investigate these in our model, or some extension of it.

\section*{Acknowledgments}

We thank S. Giddings, D. Gross, G. Horowitz, D. Kabat, D. Mateos, M. Srednicki, and B. Shraiman for comments and discussions.  This work was supported in part by NSF grants PHY05-51164 and PHY04-56556.

 \newpage


\end{document}